Diversity and abundance of the Abnormal chromosome 10 meiotic drive complex in *Zea mays*

Lisa B. Kanizay[1], Tanja Pyhäjärvi[2], Elizabeth G. Lowry[3], Matthew B. Hufford[2], Daniel G. Peterson[4], Jeffrey Ross-Ibarra[2,5], R. Kelly Dawe[1,3]

[1]Department of Plant Biology, University of Georgia, Athens, GA 30602
[2]Department of Plant Sciences, University of California, Davis, CA 95616
[3]Department of Genetics, University of Georgia, Athens, GA 30602
[4]Department of Plant & Soil Sciences and Institute for Genomics, Biocomputing & Biotechnology, Mississippi State University, Mississippi State, MS 39762
[5]Center for Population Biology and Genome Center, University of California, Davis, CA 95616

Corresponding author:
R. Kelly Dawe
2502 Miller Plant Sciences
University of Georgia
Athens, GA 30602
Phone: 706-542-1658
Fax: 706-542-1805

**Abstract**

Maize Abnormal chromosome 10 (Ab10) contains a classic meiotic drive system that exploits asymmetry of meiosis to preferentially transmit itself and other chromosomes containing specialized heterochromatic regions called knobs. The structure and diversity of the Ab10 meiotic drive haplotype is poorly understood. We developed a BAC library from an Ab10 line and used the data to develop sequence-based markers, focusing on the proximal portion of the haplotype that shows partial homology to normal chromosome 10. These molecular and additional cytological data demonstrate that two previously identified Ab10 variants (Ab10-I and Ab10-II) share a common origin. Dominant PCR markers were used with FISH to assay 160 diverse teosinte and maize landrace populations from across the Americas, resulting in the identification of a previously unknown but prevalent form of Ab10 (Ab10-III). We find that Ab10 occurs in at least 75% of teosinte populations at a mean frequency of 15%. Ab10 was also found in 13% of the maize landraces, but does not appear to be fixed in any wild or cultivated population. Quantitative analyses suggest that the abundance and distribution of Ab10 is governed by a complex combination of intrinsic fitness effects as well as extrinsic environmental variability.

## Introduction

Major deviations from Mendel's rules have been termed meiotic drive, signifying preferential transmission of chromosomes or chromosome regions to progeny (Sandler and Novitski, 1957). Although deviations in meiosis are implied by the term, the most heavily studied examples of meiotic drive affect post-meiotic events. Three well characterized examples are the *t*-haplotype in *Mus musculus*, Segregation Distorter (SD) in *Drosophila melanogaster*, and Spore Killer (Sk) in *Neurospora intermedia* (Burt and Trivers, 2006). In these examples meiotic drive is controlled by multiple loci that encode *trans*-acting factors. The loci involved are tightly linked in a haplotype shielded from recombination by genomic rearrangements and/or proximity to pericentromeric heterochromatin (Burt and Trivers, 2006; Lyttle, 1991). In all three cases the mechanism of meiotic drive involves disruption of sperm or spore sterility (Burt and Trivers, 2006; Lyttle, 1991).

In contrast, the maize Abnormal chromosome 10 (Ab10) meiotic drive system changes meiosis in a way that results in the preferential inclusion of heterochromatin-based knobs in reproductive cells (Rhoades, 1952). Ab10 contains an extended chromatin region at the end of the long arm roughly the size of the short arm of a normal chromosome 10 (Hiatt and Dawe 2003a). This region is a complex haplotype consisting of at least three knobs and several genes (Mroczek et al., 2006) (Figure 1A). The knobs are composed of two different tandem repeats, known as the 180-bp knob repeat and the 350-bp TR-1 repeat (Figure 1A). Also within the Ab10 haplotype are at least two *trans*-acting factors that independently convert the two types of knob into "neocentromeres." Moving rapidly along spindle microtubules, neocentromeres cause knobbed chromatids to segregate to the upper and lower cells of the linear tetrad (Dawe et al., 1999; Hiatt and Dawe, 2003a). Since the lower cell of the tetrad develops into the egg, this mechanism gives Ab10 (and other knobs when Ab10 is present) a segregation advantage (Rhoades, 1952). In theory, Ab10 can reach 83% transmission as a heterozygote (Buckler et al., 1999), although measured meiotic drive levels typically range from 65-80% (Hiatt and Dawe, 2003a; Rhoades and Dempsey, 1966). The degree of preferential transmission is limited by a number of factors including recombination between centromeres and knobs and the efficiency of neocentromere formation (Hiatt and Dawe, 2003a).

Although knobs are common in maize and teosinte (Albert et al., 2010), early cytological surveys suggest that Ab10 is relatively rare, segregating in only 18% of sampled maize landrace populations and 35% of teosinte populations (McClintock et al., 1981). Meiotic drive systems are often maintained at low levels in nature due to intrinsic deleterious fitness effects and/or selection for host-encoded suppressors that reduce meiotic drive (Burt and Trivers, 2006; Lyttle, 1991). Plausible explanations for the low frequency of Ab10 include a reduced male transmission of Ab10 or other fitness consequences of Ab10 in the homozygous condition (Buckler et al., 1999). In addition, there is evidence that the diversity and frequency of Ab10 has been affected by maize domestication. For example, the Ab10-I type originally described by Rhoades (Rhoades, 1942) is the only known form of Ab10 in landraces (McClintock et al., 1981), while at least two forms are known in teosinte – the Ab10-I type and a cytologically distinct form known as Ab10-II (Kato, 1976; McClintock et al., 1981; Rhoades and Dempsey, 1985). Neither Ab10-I nor Ab10-II has been observed in any modern maize inbred lines (Albert et al., 2010).

In this study we developed molecular markers for the Ab10 haplotype and used them to study the abundance and diversity of Ab10 in maize and its wild relatives. Prior data suggest that much of the Ab10 haplotype is derived from distal sections of the normal chromosome 10 (N10), but that the N10 sequences are scattered, rearranged, and mixed with unknown sequence and transposable elements (Figure 1A) (Mroczek et al., 2006; Rhoades and Dempsey, 1985). We reasoned that if we sequenced long sections of the N10-Ab10 shared region from bacterial artificial chromosome (BAC) clones we could identify sequences and/or polymorphisms unique to Ab10. This approach allowed us to identify two Ab10-specific PCR markers. In conjunction with an extensive new cytological survey, we use these markers to document the frequency of Ab10 in maize and related wild taxa. We find that Ab10 is more common in teosintes than cultivated maize landraces, and presumably absent altogether from modern inbreds. We document a previously unknown variant of Ab10 (Ab10-III) that shows a novel knob structure and appears to be more prevalent than Ab10-I or Ab10-II. Finally, an analysis of allele frequency differentiation of Ab10 across populations suggests that the factors that determine Ab10 abundance in natural and cultivated populations are complex.

## Materials and Methods

*BAC library creation and gridding onto high-density filters*

Tissue from a lab stock homozygous for the Ab10-I haplotype was used to create a BAC library (ZMMTBa). This line has an undefined genetic background. The BAC library was created in the pIndigoBAC-5 vector as described previously (Peterson et al., 2000) using the *Hind*III restriction enzyme option and the "Y" method for nuclei extraction with minor modifications (Peterson et al., 2002). BACs were individually archived in 384-well microtiter plates. Clones were gridded and fixed onto 5 nylon membranes as described previously (Magbanua et al., 2011). The average insert size of the clones was 112kb, and the library afforded 4x coverage of the maize genome.

*Overgo probe design and library hybridization*

Single copy cDNA sequences from chromosome 10L (B73 reference genome version 1) found between the marker *rps11* (distal to the R1 locus) and the end of the chromosome were identified using BLAST at PlantGDB.org. Twenty-eight single copy overgo probes were designed from the cDNA sequences and used for BAC library screening (Suppl. Table 1). Overgo probes were diluted to a working concentration of 20 µM, labeled with $^{32}$P-dCTP and $^{32}$P-dATP, pooled, and hybridized to membranes at 55°C (Magbanua et al., 2011). Of these 28, four did not hybridize to the library and 24 identified more than one BAC clone (between two and eleven colonies). In addition, several BACs hybridized to more than one probe (Suppl. Table 1). Ninety-six positive clones were handpicked, re-grown and spotted onto nylon membranes as a sub-library. The 96 colony membranes were re-probed with overlapping pools of probes to identify correspondence between colonies and probes. Eleven BACs were chosen for sequencing, and are referred to by the probe number(s) that hybridized to them in the sub-library screen. The original library coordinates for these 11 BACs are listed in supplemental Table 2.

*BAC preparation and sequencing*

The eleven BACs were purified using the Large-Construct Kit (QIAGEN). Insert size was assayed by pulsed field gel electrophoresis of BACs digested with *Not*I (Peterson et al., 2002). Intact BACs were submitted to the Georgia Genomics Facility for 454 Titanium FLX sequencing. BAC sequences were submitted to the National Center for Biotechnology Information (NCBI) as high throughput genome sequence phase 1 (HTGS-1) (see Suppl. Table 1 for GenBank numbers).

*RNA isolation and cDNA preparation*

Immature tassels were dissected from sibling plants containing Ab10 or the canonical chromosome 10 (N10). Anthers spanning pre-meiotic to mature pollen stages were frozen in liquid nitrogen and RNA was extracted using the Qiaquick RNA extraction kit (QIAGEN). cDNA was synthesized using the Mint kit (EVROGEN), then normalized with the Trimmer kit (EVROGEN). The Ab10 and N10 normalized cDNA libraries were submitted to Emory University for 100 bp paired-end Illumina sequencing. Each library was run on its own lane and reads were assembled into *de novo* cDNA contigs by the core facility at Emory. This resulted in the identification of 67,725 cDNA contigs from Ab10-I line, and and 46,498 contigs from the N10 line.

*BAC sequence assembly and analysis*

The 454 reads were assembled with both Newbler (454 Life Sciences, Roche, Branford, CT 06405) and MIRA (http://sourceforge.net/apps/mediawiki/mira-assembler/). These assemblies and remaining raw reads were then further assembled using Sequencher (Gene Codes Corporation, Ann Arbor, MI, USA). Two BACs that hybridized to the same two probes (11-12) were assembled together and treated as a single BAC. To identify putative genes, repetitive sequences were first removed using RepeatMasker (http://www.repeatmasker.org/). Masked contigs were then mapped with BLAST to the non-redundant nucleotide and protein databases at NCBI with an e-value cut off of $10^{-5}$ (Suppl. Table 2). Gene models were identified using FGenesH (SOFTBERRY) and Augustus (http://bioinf.uni-greifswald.de/augustus/).

*Mapping BACs to Ab10 haplotypes*

Nine repeat junction primers (RJ) were identified using RJPrimers: v1.0 from the unmasked BAC sequence contigs (Luce et al., 2006; You et al., 2010) (Suppl. Table 3). Six intron size polymorphism (ISP) primers were identified by comparing the 35 complete genes identified in Ab10 BACs to their homologs in the B73 reference genome (Schnable et al., 2009) (Suppl. Table 3). Introns that differed in size by at least 50 bp between Ab10 and B73 were tested as markers. PCR was used to map the BACs within the Ab10 haplotype using the combined set of 15 primers in a series of DNAs extracted from deficiency lines (Hiatt and Dawe, 2003b). PCR was performed using 10-30 ng genomic DNA in reactions containing 1x Sigma PCR buffer, 2.5 mM $MgCl_2$, 0.25 mM dNTPs, 0.25 µM primers, and 1-2 unit Sigma *Taq* polymerase. Reactions were denatured at 94°C for 5 minutes, followed by 40 cycles at 94°C for 30 seconds, 55-59.5°C for 20 seconds, 72°C for 30-60 seconds, and a final extension at 72°C for 5 minutes.

Stocks used were Ab10-I_Rhoades (referred to as Ab10-I throughout), Ab10-II_Rhoades (referred to as Ab10-II throughout), and the deficiency lines Ab10-I-Df(I), Ab10-I-Df(F), Ab10-I-Df(H), and Ab10-I-Df(K), all originally obtained from Marcus Rhoades. Additional deficiency lines Ab10-I-Df(B), Ab10-I-Df(M), and Ab10-I-Df(L) were described previously (Hiatt and Dawe, 2003b), and Ab10-II-Df(Q) and Ab10-II-Df(M) were obtained from the Maize Genetics Cooperation Stock Center, University of Illinois, Urbana, Ill. The deficiency lines were crossed to either the B73 or W23 (N10) inbreds, and the resulting heterozygotes used for scoring the markers by presence or absence.

*Teosinte and landrace PCR screen*

The dominant RJ markers D6 (Ab10-D6) and G8 (Ab10-G8) (Suppl. Table 3) were used to screen a set of 638 DNA samples derived from 135 landraces, 10 populations of the lowland teosinte taxon *Zea mays* ssp. *parviglumis* and 10 populations of the highland teosinte taxon *Zea mays* ssp. *mexicana* (Suppl. Table 4). These two teosintes represent the closest wild relatives of maize (Matsuoka et al., 2002). Samples were selected to be both geographically and genetically representative based on previously published data (Fukunaga et al., 2005; van Heerwaarden et al., 2011; Vigouroux et al., 2008). Genomic DNA was prepared using a CTAB method (Clarke, 2009; Saghaimaroof et al., 1984).

*Fluorescence* in situ *hybridization (FISH)*

Mitotic root tip chromosomes from 281 landrace individuals were analyzed by FISH as described previously (Kato et al., 2004). Mixtures of oligo probes were used to visualize the 156 bp centromeric repeat, CentC, and the two knob repeats, TR-1 (350 bp) and knob180 (180 bp). Fluorescently-labeled DNA oligos for TR-1 (20-21 nt long, four oligos) and CentC (20-23 nt long, four oligos) were obtained from Integrated DNA Technologies (http://www.idtdna.com/). The TR-1 oligos were 5' Cy3 labeled and the CentC oligos were 5' Cy5 labeled. Ten FITC labeled DNA oligo probes for knob180 were previously designed (Yu et al., 1997). Each oligo was resuspended in 2x SSC (saline sodium citrate) buffer to 100 μM. The oligos for each repeat were mixed in equimolar amounts and diluted to a final concentration of 10 μM. The 10 μM probe mixtures were used directly for FISH of prepared slides. A solution of 0.5 µl TR-1 probe mix, 0.5 µl CentC probe mix, 0.2 µl knob180 probe mix, 5 µl salmon sperm DNA (140 ng/µl), and 3.8 µl 2x SSC in 1x TE (TRIS EDTA) buffer was dropped onto slides containing the chromosomes. Slides were then denatured 5-10 minutes in a humid chamber in a boiling water bath and allowed to hybridize at room temperature for 1-3 hours. Slides were then rinsed in 2x SSC, air-dried, and mounted in 10 µl of Vectashield with DAPI (Vector Laboratories, Burlingame, CA, USA). Meiotic chromosomes were prepared as previously described (Shi and Dawe, 2006) and subjected to the same FISH protocol. Images were collected using a Zeiss Axio Imager and processed using Slidebook 5.0 software (Intelligent Imaging Innovations, Denver, CO, USA).

*Estimating the divergence of the Ab10 haplotypes*

Two regions of the Ab10 haplotype (Figure 2) were amplified and PCR products were directly sequenced from both directions from eleven different chromosomes (seven N10 and four Ab10). The two loci assayed (3471 and 8042) can be found on chromosome 10 in the B73 genome (version 2) at positions 142093638-142093985 (3471=GRMZM2G119802) and 147609512-147609977 (8042=GRMZM2G150286). For N10 controls, we used lines from the USDA Germplasm Resources Information Network (GRIN) (http://www.ars-grin.gov/npgs): Mo17 (PI 648432), CML220 (Ames 27087), B73 (PI 550473), I137TN (Ames 27116), Tzi9 (PI 506247), K55 (Ames 22754), and CI66 (PI 587148). Two isolates of Ab10-I were used, Ab10-I_Rhoades, originally collected from a site outside of Mexico City, and a second isolate (line X233F from the Maize Genetics Cooperation Stock Center) that was collected from the American Southwest. Our Ab10-II stock was obtained from Marcus Rhoades. The Ab10-III chromosome was identified in a homozygous state in the landrace Guatemala 110 (PI 490825 from the USDA Germplasm Resources Information Network).

The sequences were used to estimate phylogeny and haplotype divergence between Ab10 and N10. In order to investigate the phylogenetic relationship of N10 and Ab10 we aligned sequences in MEGA (Tamura et al., 2007) and created maximum likelihood trees with 500 bootstrap iterations, using a Jukes-Cantor nucleotide substitution model. The divergence time between haplotype groups was estimated as $T = 2L\mu/d$, where T is the divergence time in years, L the length of the sequence in bp, $\mu$ the mutation rate per bp and d the distance between haplotypes estimated using counts of fixed differences or Nei's net divergence (Nei, 1987). In all cases we assumed a mutation rate of $3 \times 10^{-8}$, similar to recent estimates from maize (Clark et al., 2005).

*Frequency of Ab10 in teosinte populations*

We used the software BayeScan 2.01 (Foll and Gaggiotti, 2008) to investigate whether Ab10 frequencies in teosinte populations deviate from neutral expectations. The method is based on a model where both locus specific and population specific components affect allele frequencies. The posterior probability of selection affecting a locus is evaluated by comparing the probabilities of models with and without selection using a Markov Chain Monte Carlo approach. We conducted the analysis for both subspecies together and separately to account for the hierarchical population structure of the two teosinte subspecies (Pyhäjärvi et al., 2012). SNPs genotyped in the same teosinte individuals (Pyhäjärvi et al., 2012) used in our Ab10 PCR screen were transformed into dominant markers by randomly treating one allele as recessive. These SNPs exhibit unusually high average expected heterozygosity ($H_E$) due to the ascertainment scheme of the markers included on the genotyping chip. To remove this effect, only 1907 SNPs that had similar $H_E$ (0.05-0.1) to the D6 and G8 Ab10 PCR markers ($H_E$ 0.09 and 0.07) were included in the analysis. Estimated heterozygosities were based on the assumption of Hardy-Weinberg equilibrium. BayesScan was run with default parameters: 20 pilot runs, a burn-in period of 50,000 iterations and 5,000 output iterations with a thinning interval of 10. Prior boundaries for the inbreeding coeffcient ($F_{IS}$) for each population were from 0 to 0.1 based on (Pyhäjärvi et al., 2012).

## Results

*Sequence of Ab10 BACs*

We created a BAC library from a line homozygous for Ab10-I and screened it with multiple probes distributed across the region shared between N10 and Ab10 (Suppl. Table 1, Figure 2). Eleven BACs were sequenced using 454 technology (two of which overlapped and were treated as one). Gene prediction programs suggested the presence of at least 43 genes. To confirm these, we sequenced the transcriptomes from an Ab10-I plant and wild type (N10) sibling and confirmed that at least 35 of the Ab10-I genes were transcribed (See Materials and Methods and Suppl. Table 2). All but one of the 35 confirmed Ab10 genes showed clear homology to known genes on the long arm of chromosome 10 in the B73 reference genome. In the single exceptional case, the homolog in B73 is present on chromosome 9 (GRMZM5G811697).

*Development of molecular markers to map Ab10 BACs*

One defining feature of Ab10 is that it does not recombine with N10 distal to the *R1* locus, making traditional mapping a challenge (Rhoades, 1942). Deletion mapping is a viable alternative since large terminal deficiencies of Ab10 can be transmitted through the female (Hiatt and Dawe, 2003b). Previous studies have used deficiencies of Ab10-I and Ab10-II to map shared genes with mutant phenotypes: *White Seedling2* (*W2*), *Opaque Endosperm7* (*O7*), *Luteus13* (*L13*), and *Striate Leaves2* (*Sr2*) (Rhoades and Dempsey, 1985; Rhoades and Dempsey, 1988). Ab10-I deficiencies have also been used to map RFLP markers (Mroczek et al., 2006). In principle the Ab10-I and Ab10-II deficiencies can be used to map any marker, as long as the marker is dominant or co-dominant, since nearly all of the Ab10 deficiencies are homozygous inviable and must be propagated as heterozygotes with N10 (Hiatt and Dawe, 2003b).

We identified at least one PCR marker per BAC (Suppl. Table 3). Among these were six intron size polymorphisms and nine repeat junction (RJ) markers (Luce et al., 2006; You et al., 2010) that could be used to differentiate Ab10 from N10. Eight deletion lines exist for Ab10-I, however only two deletion lines exist for Ab10-II, and these two have been poorly characterized (Rhoades and Dempsey, 1988). By scoring the presence or absence of the markers in the Ab10 deficiency strains we established that BAC2-3 mapped upstream of the most proximal known breakpoints of both Ab10-I and Ab10-II (Figure 2). On Ab10-I, all other mapped BACs are located between the Df(I) and Df(F) breakpoints, consistent with prior data confirming the existence of a large inversion within the Ab10 haplotype. A similar inversion exists on Ab10-II. Four BACs that correspond to the proximal side of Ab10-II (between Df(Q) and Df(M)) map to the distal side of N10, and four others from the distal region of Ab10-II map to the proximal side of N10 (Figure 2). None of the BACs, which span nearly all of the shared region, map distal to the Ab10-I-Df(L) breakpoint.

*Identification of Ab10 specific markers for population screens*

Ab10 was not observed in a prior cytological screen of 103 maize inbreds (Albert et al., 2010). We tested all of our Ab10-based primer pairs in a set of 53 inbreds (Shi et al., 2010) that includes 24 lines not assayed in Albert et al. (2010). Two of the dominant RJ PCR markers, Ab10-D6 ("D6") and Ab10-G8 ("G8"), were not found in any of these lines, suggesting they are unique to Ab10. To further test their reliability, we carried out a larger test screen that combined PCR with fluorescent *in situ* hybridization (FISH) on a total of 150 individuals from 19 maize landraces (Suppl. Table 4). We chose landrace accessions based on their ready availability in the National Plant Germplasm System (http://www.ars-grin.gov/npgs). These data revealed that 11 of the 19 landraces segregate for Ab10. A total of 34 individuals contained Ab10 as scored by both PCR and FISH. The PCR and FISH results were entirely concordant, although many Ab10 chromosomes carried only one of the two markers (either D6 or D8) (Table 1, Suppl. Table 4). Another 116 individuals tested negative for Ab10 by both FISH and PCR (Suppl. Table 4).

*Discovery of a new cytological variant, Ab10-III*

While performing FISH on our landraces, we observed that four landraces were segregating Ab10-I, and unexpectedly, that nine landraces carried a new cytological variation of the Ab10 haplotype that we term Ab10-III. Ab10-II was not observed in landraces (consistent with prior conclusions (McClintock et al., 1981). Ab10-III appears to be similar to Ab10-I in the central domain, but contains a second TR-1-rich domain appended to the end of the main knob

(Figure 1B). The two dominant markers D6 and G8 appear to segregate independently of this cytological polymorphism (Table 1, Suppl. Table 4). While the Ab10-I_Rhoades haplotype used to create the BAC library carries both markers, the three other landraces with chromosomes that appear to be similar to Ab10-I scored positive for G8 but not D6. Ab10-III was observed to have D6, G8, or both D6 and G8 depending on the isolate (Suppl. Table 4).

*Estimating divergence time of haplotypes*

The relationship of Ab10 and N10 was examined by sequencing two loci in the shared region, one from BAC 2/3 (3741) and another from BAC 16 (8042) (Figure 2). Gene trees of the two loci revealed an unexpected grouping pattern (Figure 3), in which Ab10-I and Ab10-II form one group while Ab10-III consistently groups with N10 haplotypes. The Ab10 (-I and -II) and N10 (+Ab10-III) groups differ at 11 fixed SNPs within the 357-bp locus 3741 and 8 SNPs within the 510-bp locus 8042. Estimates of divergence time based on these fixed SNPs suggest the groups split approximately 365,000 years ago. Estimates based on net pairwise nucleotide divergence (Nei, 1987), suggest the split occurred 535,000 years ago for locus 3741 and 377,000 years ago for locus 8042. These data suggest that the shared region of Ab10 is not markedly older than other regions of the maize genome as estimated from nucleotide diversity (Doebley and Iltis, 1980; Ross-Ibarra et al., 2009).

*Ab10 abundance and allele frequencies in teosinte and landraces*

Using the dominant D6 and G8 markers, we assessed the frequency of Ab10 in ten natural populations of *Zea mays* ssp. *parviglumis*, and ten natural populations of *Zea mays* ssp. *mexicana* at a sample size of 12 individuals each (for a total of 240 individuals). We also

16examined a total of 571 landrace individuals with either FISH or PCR from multiple sources and population sizes (including the 150 discussed above; Suppl. Table 4, Figure 4). These data revealed that Ab10 is found in 14% of *parviglumis* (17 positive individuals from 7 populations), 16% of *mexicana* (19 positive individuals from 8 populations), and 13% of maize landraces (68 positive individuals). The two markers differ in frequency among the three subspecies, with D6 being more prevalent in *mexicana* than either *parviglumis* or maize landraces (Table 1, Suppl. Table 4).

While the frequency of Ab10 varied among populations, we found no evidence for stronger differentiation than expected based on a panel of nearly 2000 other randomly chosen SNPs from the *Zea mays* genome (Figure 5A). There is an interesting negative correlation between Ab10 and altitude in both teosinte subspecies (Figure 5B, r=-0.45 for each), however this correlation is not significant, and in general, Ab10 shows weaker differentiation in teosinte than a subset of SNPs previously shown to correlate with environmental variables (Pyhäjärvi et al., 2012). In contrast, the negative correlation with altitude is statistically significant for maize landraces (Figure 5B; $p = 0.027$, $r = -0.39$). These data suggest that Ab10 is widely distributed across the species' range with some evidence for higher frequencies at low altitudes.

## Discussion

Here we provide the first molecular characterization of the proximal section of maize Abnormal chromosome 10. These data allowed us to better characterize the two known variants of Ab10 (Ab10-I and Ab10-II) and to identify a previously unknown cytological variant called Ab10-III. We show that Ab10-I and Ab10-II are similar in overall structure and ancestry, and that the strongest homology between N10 and Ab10 lies in the regions proximal to the major knob. The "distal tip" is not derived from modern chromosome 10 and its origin remains obscure, yet is it known to contain at least one of the key meiotic drive functions (Hiatt and Dawe 2003a).

Using two Ab10-specific markers in conjunction with FISH we scored a total of 769 *Zea mays* (maize and teosinte) individuals for Ab10. Our data demonstrate that Ab10 is more common and more diverse than originally thought. Figure 6 compares our Ab10 frequency data to the prior data accumulated by McClintock and coworkers (McClintock et al., 1981). The same trends are apparent in both data sets: Ab10 is more common in teosinte populations than maize landrace populations, and Ab10 is rarely observed in more than 40% of the individuals in any given population. The overall frequency of Ab10 in current germplasm reserves of maize landraces is roughly 13%. Among the 281 landrace individuals scored by FISH during the course of this study, we observed Ab10-I infrequently, never observed Ab10-II, and observed Ab10-III most frequently (Suppl. Table 4 and data not shown). We presume that Ab10-II is a common variant in teosintes (Kato, 1976; McClintock et al., 1981), however in this study we did not conduct FISH assays in teosinte lines.

We were surprised to find that the central shared domain of Ab10 is quite variable in molecular structure. Sequence analysis of two loci within four Ab10 haplotypes (two Ab10-I haplotypes, Ab10-II and Ab10-III) suggest a complex ancestry; while Ab10-I and Ab10-II are clearly related, Ab10-III appears to have either arisen independently on an N10 background or exchanged material with N10 haplotypes via gene conversion (Figure 3). We also show that the two PCR markers, D6 and G8, cannot be used to differentiate between Ab10-I, Ab10-II or Ab10-III, suggesting the haplotypes have recombined with each other (Suppl. Table 4). Across subspecies, G8 is more prominent in landraces and *parviglumis* compared to D6, however D6 is more prevalent than G8 in *mexicana* (Table 1).

At overall frequencies of 12-15%, different variants of Ab10 must regularly come into contact with each other, where they presumably recombine in the proximal regions, and compete with each other for preferential transmission. As knobs on other chromosomes compete for transmission based on their size (Kikudome, 1959), it follows that head-to-head competition between different Ab10 variants will favor haplotypes with larger knobs. In fact the knobs on Ab10 are generally the largest in the genome (Kato, 1976; McClintock et al., 1981), even though they compete with knobless chromosomes in most populations. In accordance with this expectation, Ab10-III differs from the other haplotypes primarily by the amount and types of knob repeats. The two major knob repeats have independent neocentromere activities, with TR-1 being able to move poleward faster than the knob180 repeat (Hiatt et al., 2002). The additional TR-1 knob found on Ab10-III may provide this haplotype with an advantage over the other haplotypes that has allowed it to become the most prevalent form of Ab10 in landraces.

Drive alone should cause Ab10 to quickly fix across populations (Buckler et al., 1999). The fact that Ab10 is observed at intermediate frequencies thus implies the action of additional intrinsic (homozygous disadvantage, pollen inviability, etc.) or extrinsic (environmental effects, gene flow, population colonization history) factors on Ab10. Homozygous Ab10 plants will grow in cultivation but their fitness in natural populations is not known. Intrinsic limitations on homozygosity may help explain why Ab10 is widespread but rarely rises above 40% in any given population, as well as the fact that it is absent from inbred lines. But while the frequency of Ab10 is not identical across populations – as might be predicted under a simple equilibrium model – it does not show stronger differentiation (e.g. higher Fst) than other markers in the genome (Figure 5A). The primary extrinsic limitation on Ab10 frequency may be factors correlated with altitude ((Buckler et al., 1999); Figure 4). For example, selection for faster growth at higher altitudes may select against larger genome size (Poggio et al., 1998) and thus also Ab10 and the multitude of associated heterochromatic knobs that it drives to high frequency (Buckler et al., 1999).

**Funding**

This work was supported by a grant from the National Science Foundation (NSF-0951091).

**Acknowledgements**

We would like to thank Lauren Sagara for help with DNA extractions. Xueyan Shan, Melanie Smith, Calla Kingery, and Zenaida Magbanua provided help with BAC library construction and probing. We are grateful to Ryan Weil for helping with cDNA sequence assembly, Saravanaraj Ayyampalayam for helping with BAC sequence assembly, and Michael McKain for general bioinformatics guidance.

**References**

Albert PS, Gao Z, Danilova TV, Birchler JA, 2010. Diversity of Chromosomal Karyotypes in Maize and Its Relatives. Cytogenetic and Genome Research 129:6-16. doi: 10.1159/000314342.

Buckler ES, Phelps-Durr TL, Buckler CSK, Dawe RK, Doebley JF, Holtsford TP, 1999. Meiotic drive of chromosomal knobs reshaped the maize genome. Genetics 153:415-426.

Burt A, Trivers R, 2006. Genes in conflict: the biology of selfish genetic elements. Cambridge, Massachusetts: The Belknap Press of Harvard University Press.

Clark RM, Tavare S, Doebley J, 2005. Estimating a nucleotide substitution rate for maize from polymorphism at a major domestication locus. Molecular Biology and Evolution 22:2304-2312. doi: 10.1093/molbev/msi228.

Clarke JD, 2009. Cetyltrimethyl Ammonium Bromide (CTAB) DNA Miniprep for Plant DNA Isolation. Cold Spring Harbor Protocols 4. doi: 10.1101/pdb.prot5177.

Dawe RK, Reed LM, Yu HG, Muszynski MG, Hiatt EN, 1999. A maize homolog of mammalian CENPC is a constitutive component of the inner kinetochore. Plant Cell 11:1227-1238.

Doebley JF, Iltis HH, 1980. Taxonomy of Zea (gramineae) .1. a subgeneric classification with key to taxa. American Journal of Botany 67:982-993. doi: 10.2307/2442441.

Foll M, Gaggiotti O, 2008. A Genome-Scan Method to Identify Selected Loci Appropriate for Both Dominant and Codominant Markers: A Bayesian Perspective. Genetics 180:977-993. doi: 10.1534/genetics.108.092221.

Fukunaga K, Hill J, Vigouroux Y, Matsuoka Y, Sanchez J, Liu KJ, Buckler ES, Doebley J, 2005. Genetic diversity and population structure of teosinte. Genetics 169:2241-2254. doi: 10.1534/genetics.104.031393.

Hiatt EN, Dawe RK, 2003a. Four loci on abnormal chromosome 10 contribute to meiotic drive in maize. Genetics 164:699-709.

Hiatt EN, Dawe RK, 2003b. The meiotic drive system on maize abnormal chromosome 10 contains few essential genes. Genetica 117:67-76.

Hiatt EN, Kentner EK, Dawe RK, 2002. Independently regulated neocentromere activity of two classes of tandem repeat arrays. Plant Cell 14:407-420. doi: 10.1105/tpc.010373.

Kato A, Lamb J, Birchler J, 2004. Chromosome painting using repetitive DNA sequences as probes for somatic chromosome identification in maize. Proc. Nat. Acad. Sci. USA 101:13554-13559.

Kato YTA, 1976. Cytological studies of maize (Zea mays L.) and teosinte (Zea mexicana Shrader Kuntze) in relation to thier origin and evolution. Massachusettes Agricultural Experiment Station Bulletin 635:1-185.

Kikudome G, 1959. Studies on the Phenomenon of Preferential Segregation in Maize. Genetics 44:815-831.

Luce AC, Sharma A, Mollere OSB, Wolfgruber TK, Nagaki K, Jiang JM, Presting GG, Dawe RK, 2006. Precise centromere mapping using a combination of repeat junction markers and chromatin immunoprecipitation-polymerase chain reaction. Genetics 174:1057-1061. doi: 10.1534/genetics.106.060467.

Lyttle TW, 1991. Segregation distorters. Annual Review of Genetics 25:511-557. doi: 10.1146/annurev.ge.25.120191.002455.

Magbanua ZV, Ozkan S, Bartlett BD, Chouvarine P, Saski CA, Liston A, Cronn RC, Nelson CD, Peterson DG, 2011. Adventures in the Enormous: A 1.8 Million Clone BAC Library for the 21.7 Gb Genome of Loblolly Pine. Plos One 6. doi: e16214 10.1371/journal.pone.0016214.

Matsuoka Y, Vigouroux Y, Goodman MM, Sanchez GJ, Buckler E, Doebley J, 2002. A single domestication for maize shown by multilocus microsatellite genotyping. Proceedings of the National Academy of Sciences of the United States of America 99. doi: 10.1073/pnas.052125199.

McClintock B, Yamakake T, Blumenschein A, 1981. Chromosome Constitution of Races of Maize: its significance in the interpretation of relationships between races and varieties in the Americas. Chapingo, Mexico: Colegio de Postgraduados.

Mroczek RJ, Melo JR, Luce AC, Hiatt EN, Dawe RK, 2006. The maize Ab 10 meiotic drive system maps to supernumerary sequences in a large complex haplotype. Genetics 174:145-154. doi: 10.1534/genetics.105.048322.

Nei M, 1987. Molecular evolutionary genetics. Molecular evolutionary genetics:i.

Peterson DG, Tomkins JP, Frisch DA, Wing RA, Paterson AH, 2000. Construction of plant bacterial artificial chromosome (BAC) libraries: An illustrated guide. Journal of Agricultural Genomics 5.

Peterson DG, Tomkins JP, Frisch DA, Wing RA, Paterson AH, 2002. Construction of plant bacterial artificial chromosome (BAC) libraries: An illustrated guide. 2nd ed http://www.mgel.msstate.edu/pubs/bacman2.pdf.

Poggio L, Rosato M, Chiavarino AM, Naranjo CA, 1998. Genome size and environmental correlations in maize (Zea mays ssp. mays, Poaceae). Annals of Botany 82. doi: 10.1006/anbo.1998.0757.

Pyhäjärvi T, Hufford MB, Mezmouk S, Ross-Ibarra J, 2012. Complex patterns of local adaptation in teosinte. arXiv:12080634v1 http://arxiv.org/abs/1208.0634.

Rhoades M, 1942. Preferential Segregation in Maize. Genetics 27:395-407.

Rhoades M, 1952. Preferential Segreagation in Maize. Ames, IA: Iowa State College Press.

Rhoades M, Dempsey E, 1985. Structural heterogeneity of chromosome 10 in races of maize and teosinte. New York: Alan R. Liss, Inc.

Rhoades M, Dempsey E, 1988. Structure of K10-II chromosome and comparison with K10-I. 62:33-34.

Rhoades MM, Dempsey E, 1966. Effect of abnormal chromosome 10 on preferential segregation and crossing over in maize. Genetics 53:989-&.

Ross-Ibarra J, Tenaillon M, Gaut BS, 2009. Historical Divergence and Gene Flow in the Genus Zea. Genetics 181:1397-1409. doi: 10.1534/genetics.108.097238.

Saghaimaroof MA, Soliman KM, Jorgensen RA, Allard RW, 1984. Ribosomal dna spacer-length polymorphisms in barley - mendelian inheritance, chromosomal location, and population-dynamics. Proceedings of the National Academy of Sciences of the United States of America-Biological Sciences 81:8014-8018. doi: 10.1073/pnas.81.24.8014.

Sandler L, Novitski E, 1957. Meiotic drive as an evolutionary force. American Naturalist 91:105-110.

Schnable PS, Ware D, Fulton RS, Stein JC, Wei F, Pasternak S, Liang C, Zhang J, Fulton L, Graves TA, Minx P, Reily AD, Courtney L, Kruchowski SS, Tomlinson C, Strong C, Delehaunty K, Fronick C, Courtney B, Rock SM, Belter E, Du F, Kim K, Abbott RM, Cotton M, Levy A, Marchetto P, Ochoa K, Jackson SM, Gillam B, Chen W, Yan L, Higginbotham J, Cardenas M, Waligorski J, Applebaum E, Phelps L, Falcone J, Kanchi K, Thane T, Scimone A, Thane N, Henke J, Wang T, Ruppert J, Shah N, Rotter K,

Hodges J, Ingenthron E, Cordes M, Kohlberg S, Sgro J, Delgado B, Mead K, Chinwalla A, Leonard S, Crouse K, Collura K, Kudrna D, Currie J, He R, Angelova A, Rajasekar S, Mueller T, Lomeli R, Scara G, Ko A, Delaney K, Wissotski M, Lopez G, Campos D, Braidotti M, Ashley E, Golser W, Kim H, Lee S, Lin J, Dujmic Z, Kim W, Talag J, Zuccolo A, Fan C, Sebastian A, Kramer M, Spiegel L, Nascimento L, Zutavern T, Miller B, Ambroise C, Muller S, Spooner W, Narechania A, Ren L, Wei S, Kumari S, Faga B, Levy MJ, McMahan L, Van Buren P, Vaughn MW, Ying K, Yeh C-T, Emrich SJ, Jia Y, Kalyanaraman A, Hsia A-P, Barbazuk WB, Baucom RS, Brutnell TP, Carpita NC, Chaparro C, Chia J-M, Deragon J-M, Estill JC, Fu Y, Jeddeloh JA, Han Y, Lee H, Li P, Lisch DR, Liu S, Liu Z, Nagel DH, McCann MC, SanMiguel P, Myers AM, Nettleton D, Nguyen J, Penning BW, Ponnala L, Schneider KL, Schwartz DC, Sharma A, Soderlund C, Springer NM, Sun Q, Wang H, Waterman M, Westerman R, Wolfgruber TK, Yang L, Yu Y, Zhang L, Zhou S, Zhu Q, Bennetzen JL, Dawe RK, Jiang J, Jiang N, Presting GG, Wessler SR, Aluru S, Martienssen RA, Clifton SW, McCombie WR, Wing RA, Wilson RK, 2009. The B73 Maize Genome: Complexity, Diversity, and Dynamics. Science 326:1112-1115. doi: 10.1126/science.1178534.

Shi J, Dawe RK, 2006. Partitioning of the maize epigenome by the number of methyl groups on histone H3 lysines 9 and 27. Genetics 173:1571-1583. doi: 10.1534/genetics.106.056853.

Shi J, Wolf SE, Burke JM, Presting GG, Ross-Ibarra J, Dawe RK, 2010. Widespread Gene Conversion in Centromere Cores. PLoS Biology 8:e1000327. doi: 10.1371/journal.pbio.1000327.

Tamura K, Dudley J, Nei M, Kumar S, 2007. MEGA4: Molecular evolutionary genetics analysis (MEGA) software version 4.0. Molecular Biology and Evolution 24:1596-1599. doi: 10.1093/molbev/msm092.

van Heerwaarden J, Doebley J, Briggs WH, Glaubitz JC, Goodman MM, Sanchez Gonzalez JdJ, Ross-Ibarra J, 2011. Genetic signals of origin, spread, and introgression in a large sample of maize landraces. Proceedings of the National Academy of Sciences of the United States of America 108:1088-1092. doi: 10.1073/pnas.1013011108.

Vigouroux Y, Glaubitz JC, Matsuoka Y, Goodman MM, Jesus Sanchez G, Doebley J, 2008. Population structure and genetic diversity of new world maize races assessed by DNA microsatellites. American Journal of Botany 95:1240-1253. doi: 10.3732/ajb.0800097.

You FM, Wanjugi H, Huo N, Lazo GR, Luo M-C, Anderson OD, Dvorak J, Gu YQ, 2010. RJPrimers: unique transposable element insertion junction discovery and PCR primer design for marker development. Nucleic Acids Research 38:W313-W320. doi: 10.1093/nar/gkq425.

Yu HG, Hiatt EN, Chan A, Sweeney M, Dawe RK, 1997. Neocentromere-mediated chromosome movement in maize. Journal of Cell Biology 139:831-840.

**Figure legends**

**Figure 1**. Structural variants of maize chromosome 10. (**A**) Schematic of the distal portion of the long arm of normal 10 (N10), Abnormal 10 type 1 (Ab10-I), type 2 (Ab10-II), and type 3 (Ab10-III). N10 possesses no knobs and the Ab10s contain different amounts of TR-1 repeat (red) and knob 180 repeat (green). The subdomains of Ab10 are the TR-1 region, shared region, 180 bp knob, and distal tip. There is a major inversion in the shared region that includes the *W2*, *O7*, *L13* and *Sr2* loci (shown in grey for Ab10-III because it is assumed but has not been verified). (**B**) Chromosome 10 karyotypes. From left to right, FISH images from root tip spreads show N10, Ab10-I, Ab10-II, and Ab10-III. Cartoon depictions are shown below. A pachytene image from a male meiocyte shows the Ab10-III haplotype in greater detail. Images show DNA (grey), the centromere repeat CentC (yellow), TR-1 (red), and knob180 (green).

**Figure 2**. Positions of probes on N10 and corresponding BACs on Ab10-I and Ab10-II. The relative positions of 28 probes used to probe the Ab10 BAC library are shown on N10. The positions of known gene markers (*rsp11*, *gln1*, etc), the loci used for molecular dating (3741 and 8042, blue), and the estimated position of the two dominant markers D6 and G8 (purple) are also shown. The mapped locations of the Ab10 BACs are shown between deficiency breakpoints ((Df(B), Df(C), etc). The general maps of Ab10-I and Ab10-II (including marker positions) are adapted from previous work (Mroczek et al., 2006), with TR-1 shown in red and knob 180 shown in green.

**Figure 3**: Maximum likelihood phylogeny of N10 inbreds and homozygous Ab10 stocks that were used for dating. This tree was created with sequence data from locus 3741. The phylogeny

for locus 8042 (not shown) gave the same result, where Ab10-I and Ab10-II grouped together while Ab10-III grouped with the N10s. K10L2 is an N10 variant described by McClintock et al. (1981) as having a knob on its long arm.

**Figure 4**: Map of screened populations showing Ab10 frequency for each population. Locations of landrace (blue), *mexicana* (red), and *parviglumis* (green) populations in North and South America (A). Zoomed in views of Mexico (B) and South America (C) are shown. The frequency of Ab10 is shown as the shaded area in each pie chart.

**Figure 5**. Population structure and clinal variation at Ab10. **(A)** BayesScan analysis of $F_{ST}$ for PCR markers G8 and D6 in teosinte populations. Shown are the distributions of $F_{ST}$ (y axis) and the $\log_{10}$ ratio of the posterior probability of a model including selection (x axis). Higher values of the log posterior odds indicate stronger evidence for selection. Data are from 1907 random SNPs (Pyhäjärvi et al., 2012) (transparent black) and Ab10 (red). For comparison, green points show SNPs strongly associated (top 1%) with environmental variables (Pyhäjärvi et al., 2012). The analysis was done separately for *parviglumis, mexicana* and for the two combined. Vertical lines indicate a false discovery rate threshold of 5%. The horizontal grey bar indicates 5-95% quantiles of $F_{ST}$. **(B)** Regression of Ab10 frequency against elevation for each subspecies. Data for maize are based on accessions pooled into 100 m altitudinal bins.

**Figure 6**. Ab10 frequency comparison of current study and study by McClintock et al. (1981). Values shown are percentage of positives out of total sampled. Teosinte subspecies designations were not known at the time of the McClintock et al. (1981) study.

**Table 1**. Summary of Ab10 allele frequencies

| Subspecies | Populations sampled | Ab10 positive populations | Individuals sampled | Total with PCR markers | D6 only | G8 only | D6 and G8 |
|---|---|---|---|---|---|---|---|
| ssp. *Mexicana* | 10 | 8 | 120 | 19 | 13 | 5 | 1 |
| ssp. *parviglumis* | 10 | 7 | 120 | 17 | 7 | 10 | 0 |
| Maize landraces | 140 | 31 | 529 | 41[1] | 7 | 25 | 9 |

[1]There were an additional 27 landrace individuals that were Ab10-positive by FISH but were not scored by PCR.

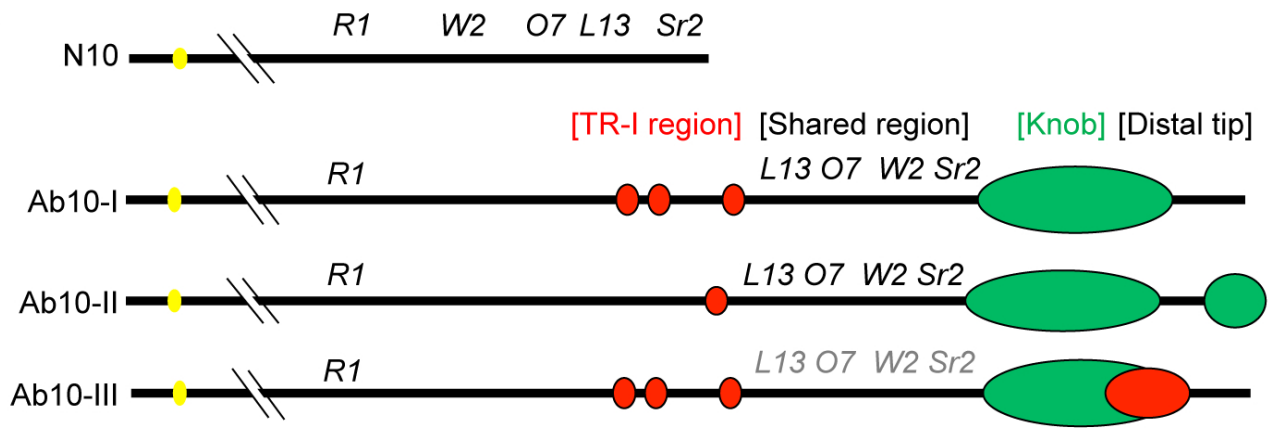

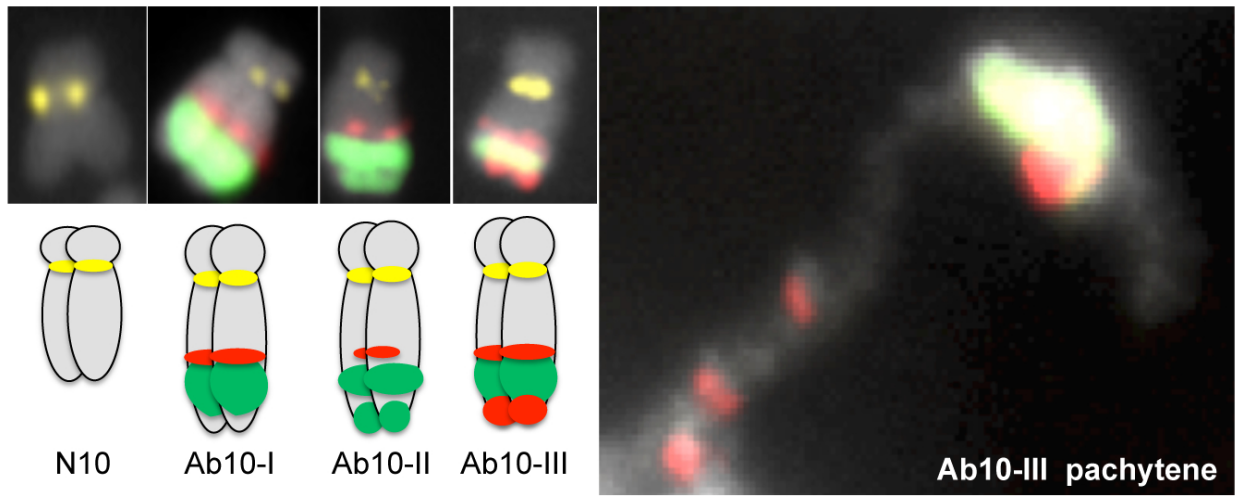

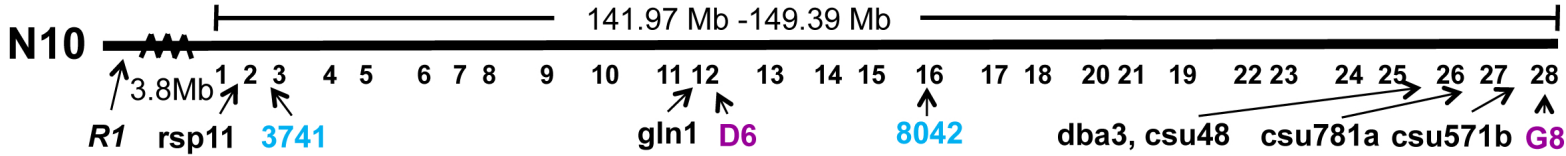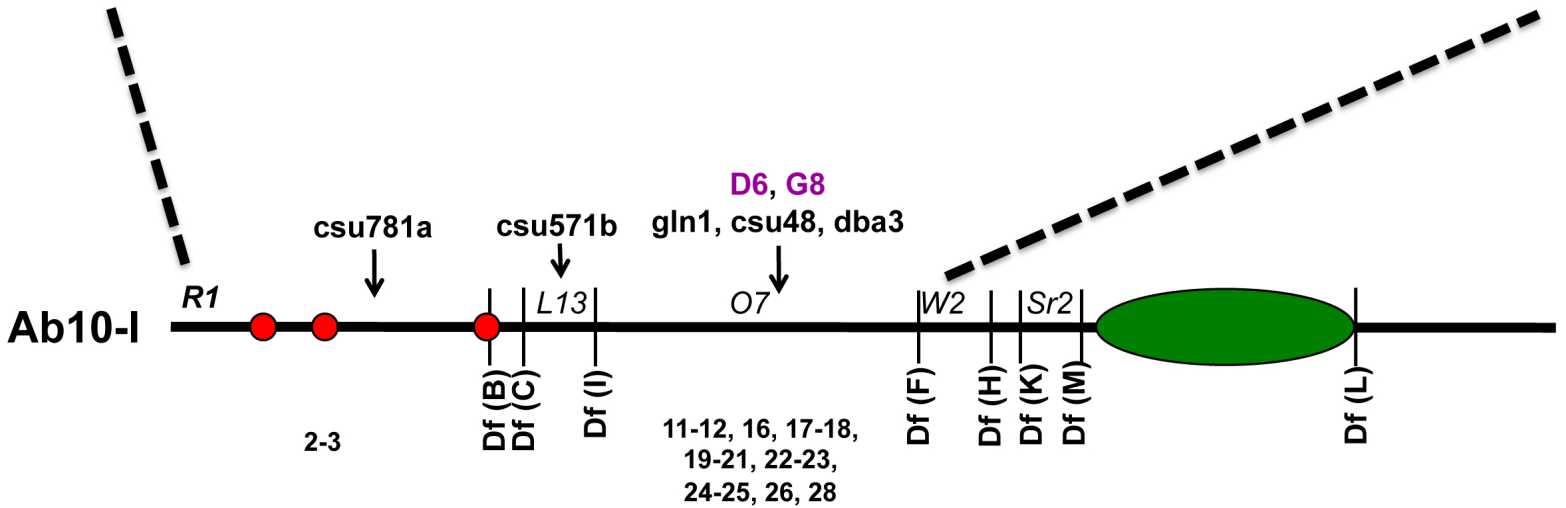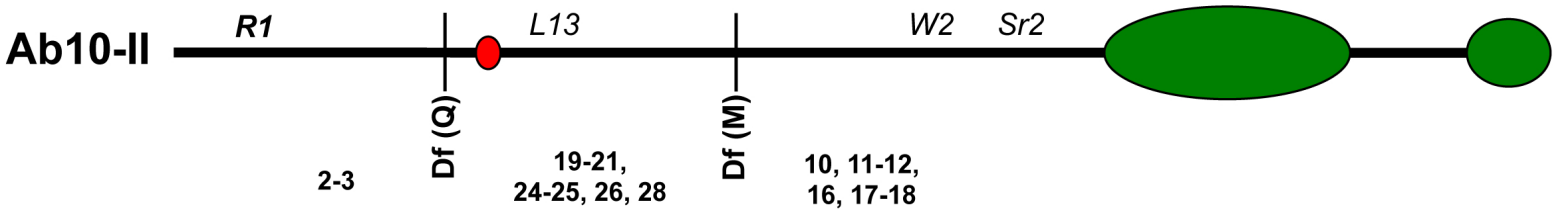

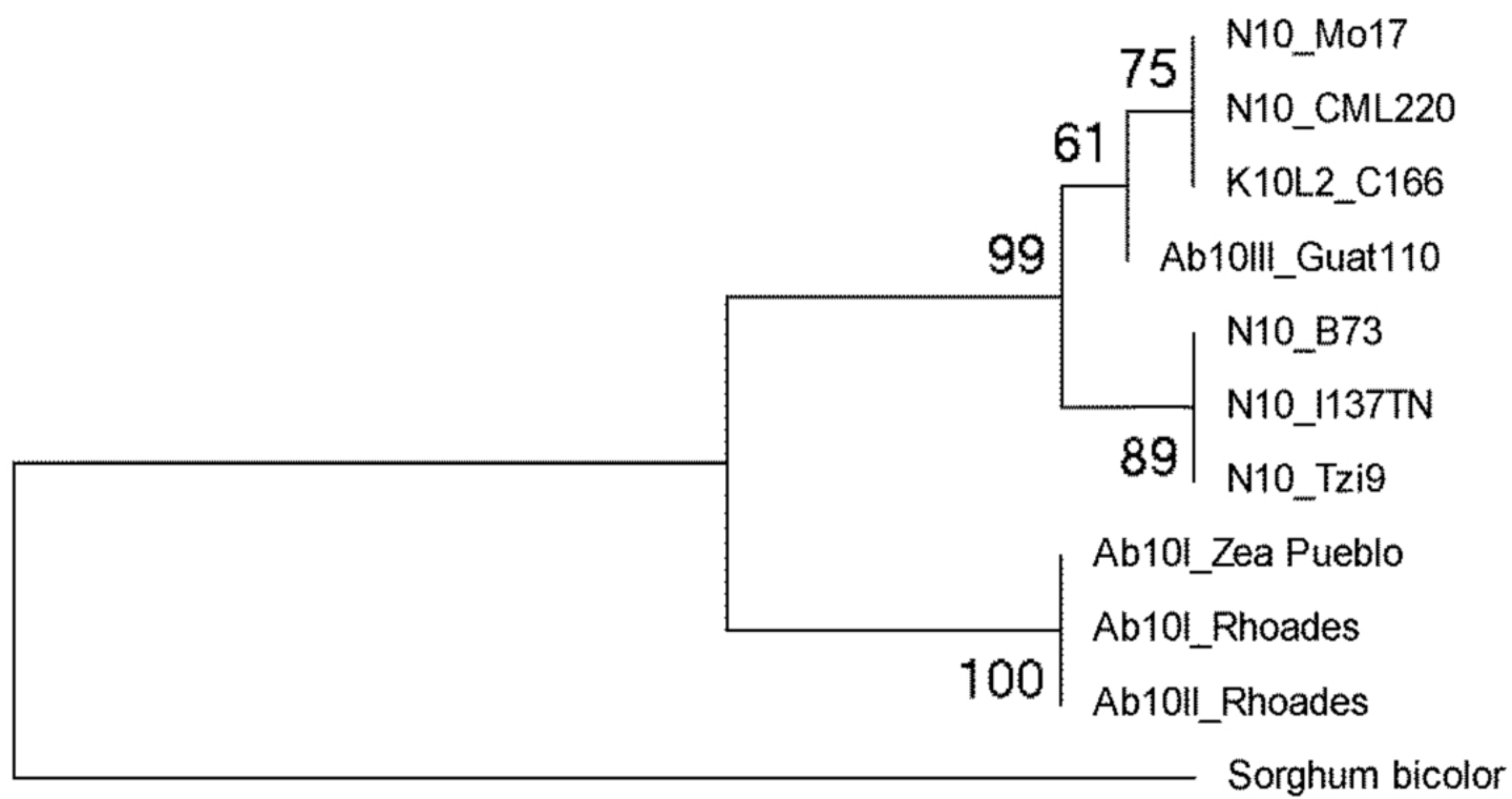

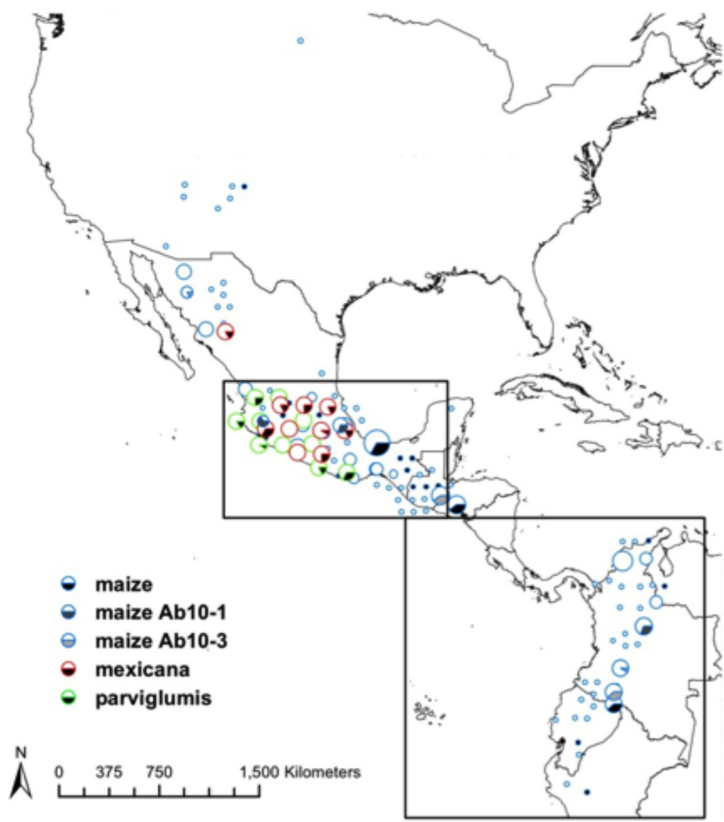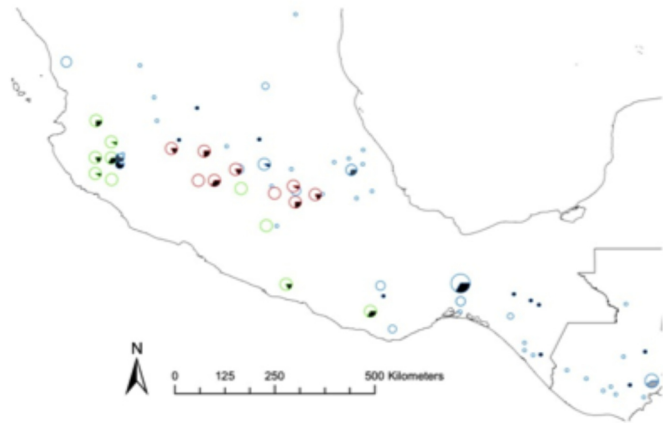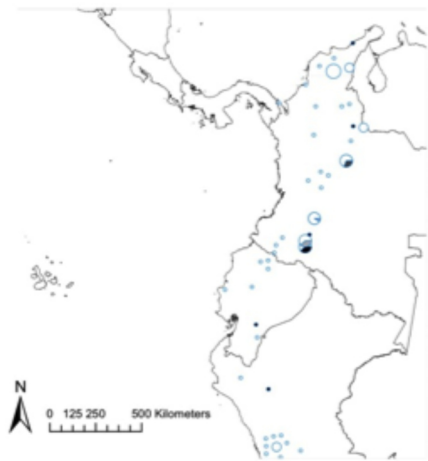

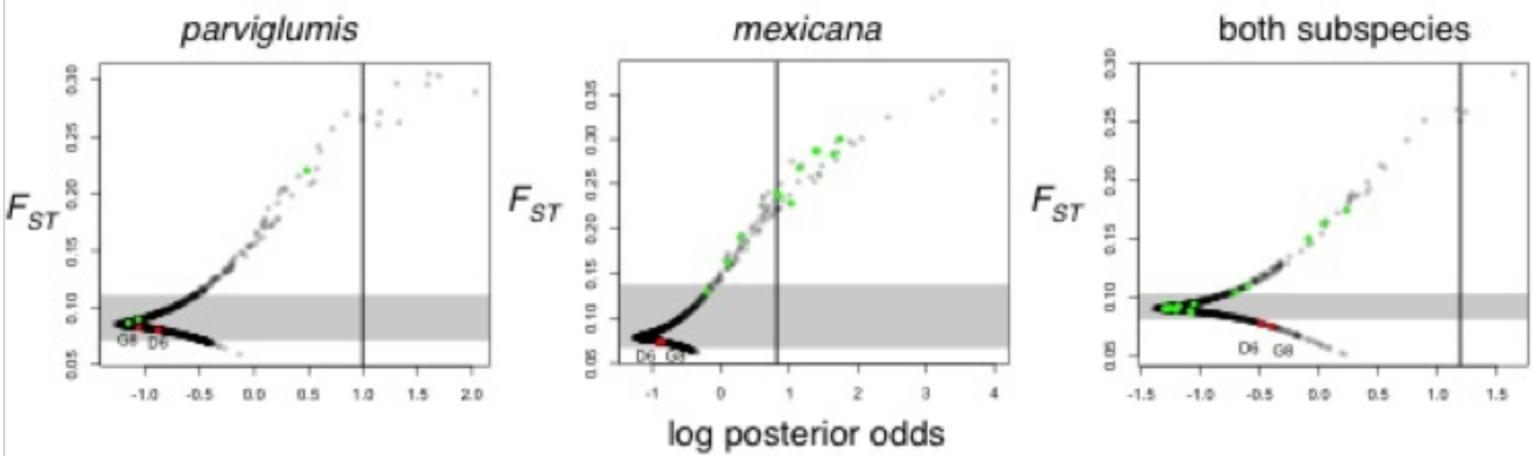

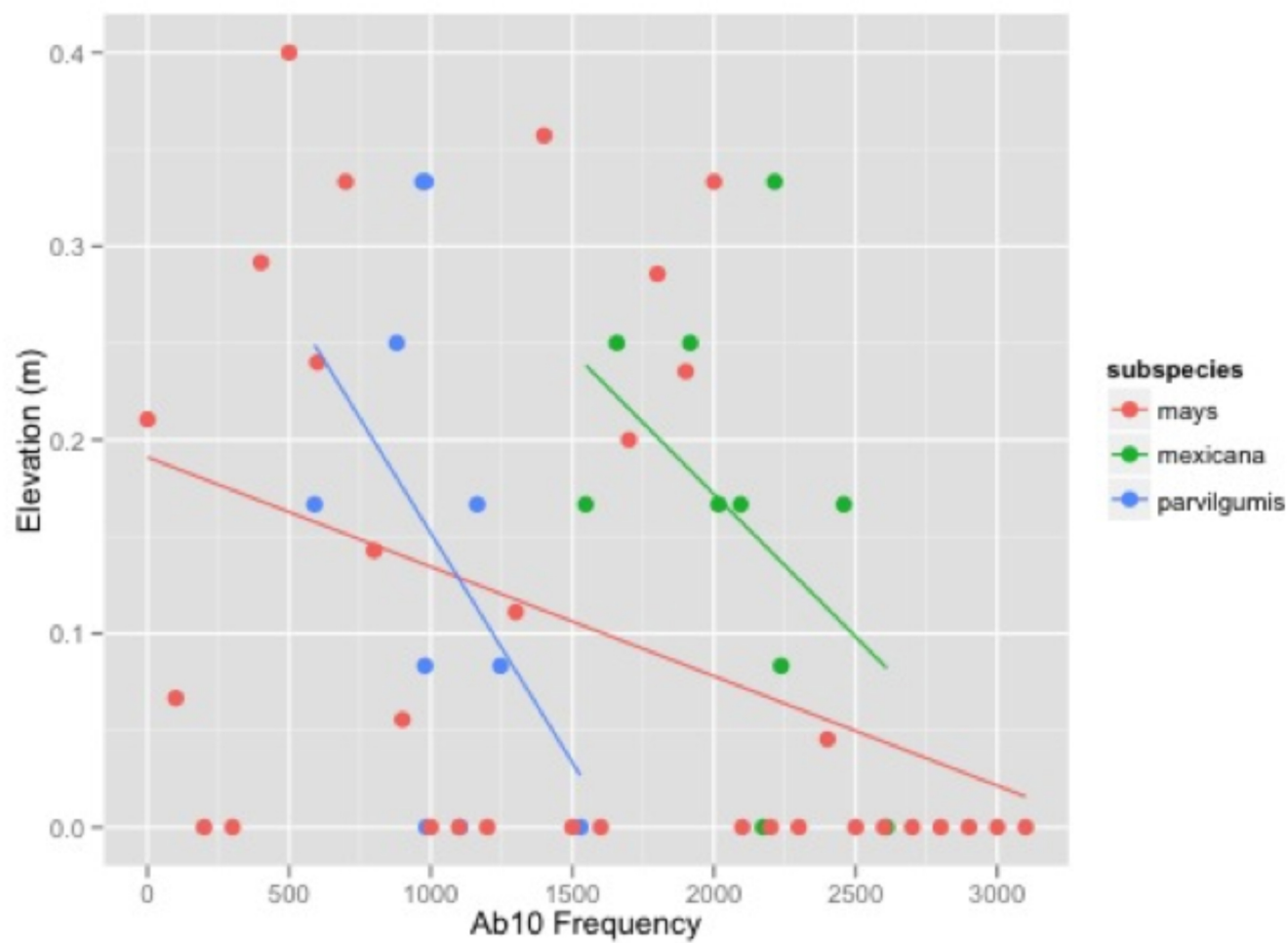

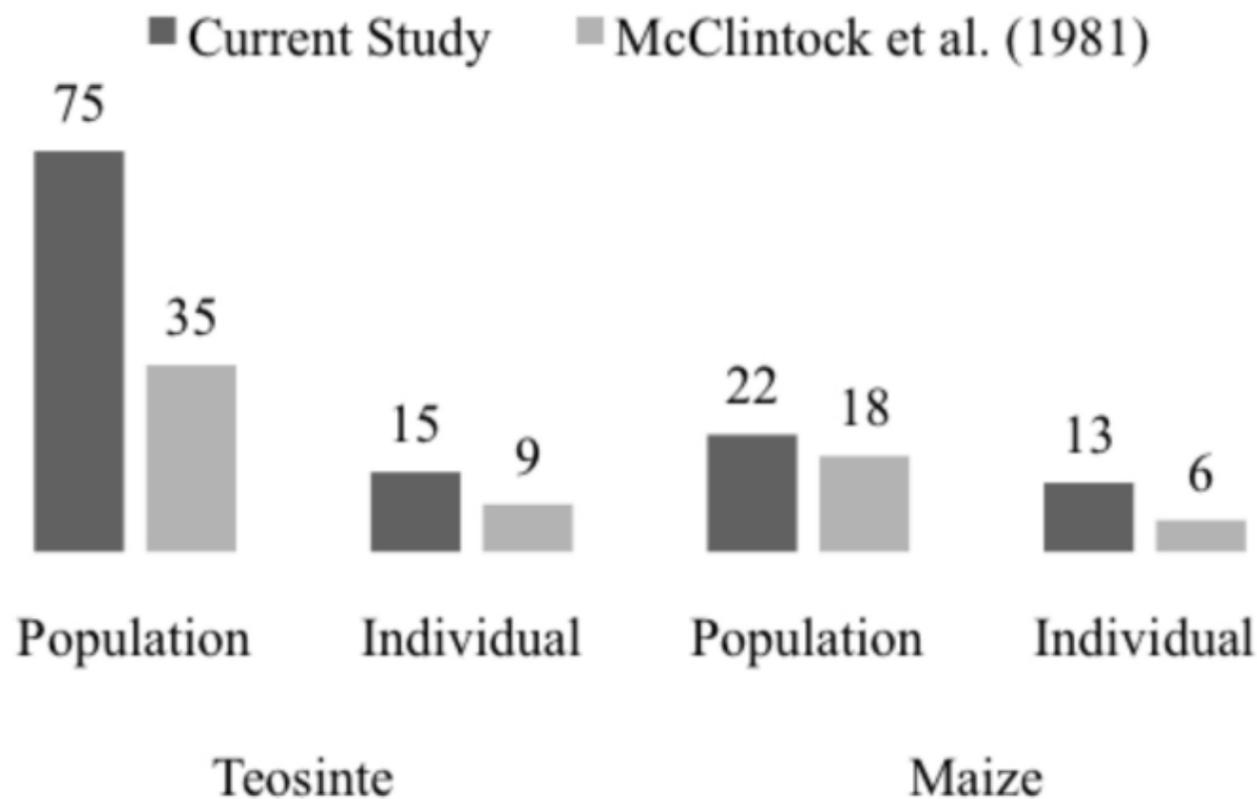